\documentclass[
twocolumn,showpacs,superscriptaddress,nofootinbib,
amssymb,a4paper
]{revtex4-1}
\usepackage[dvipdfmx]{graphicx}
\usepackage{ulem}
\usepackage{color}
\usepackage{bm}

\def \Dc{{\cal D}}

\begin{document}
%
%\setlength{\baselineskip}{16pt}
%
%---------------------------------------------------------------------------
%
\title{
Chiral effective theory of diquarks and the $\bm{U_A(1)}$ anomaly
}
%
%----------------------------------
\author{
Masayasu Harada
}
\affiliation{Department of Physics, Nagoya University, Nagoya, 464-8602, Japan
}
\author{
Yan-Rui Liu
}
\affiliation{School of Physics, Shandong University, Jinan 250100, China
}
\author{
Makoto Oka
}
\email[]{oka@post.j-parc.jp}\thanks{corresponding author}%
\affiliation{Advanced Science Research Center, Japan Atomic Energy Agency, 
Tokai, Ibaraki 319-1195, Japan
}
\affiliation{Nishina Center for Accelerator-Based Science, RIKEN, Wako 351-0198, Japan}
%
%----------------------------------
%
%
%----------------------------------
\author{
Kei Suzuki
}
\affiliation{Advanced Science Research Center, Japan Atomic Energy Agency, 
Tokai, Ibaraki 319-1195, Japan
}
%

%----------------------------------
%
% @@ =======================================================================
%
\date{\today}

\begin{abstract}
The diquark is a strongly correlated quark pair that plays an important role 
in hadrons and hadronic matter.
In order to treat the diquark as a building block of hadrons, we formulate 
an effective theory of diquark fields with $SU(3)_R\times SU(3)_L$ chiral symmetry.
We concentrate on the scalar ($0^+$) and pseudoscalar ($0^-$) diquarks and
construct a linear-sigma-model Lagrangian.
It is found that the effective Lagrangian contains a new type of chirally symmetric meson-diquark-diquark
coupling that breaks axial $U_A(1)$ symmetry.
We discuss consequences of the $U_A(1)$ anomaly term to the diquark masses as well as
to the singly heavy baryon spectrum, which is directly related to the diquark spectrum.
%We find the inverse ordering of the negative parity diquarks.
We find an inverse mass ordering between strange and nonstrange diquarks.
The parameters of the effective theory can be determined by the help of lattice QCD
calculations of diquarks and also from the mass spectrum of the singly heavy baryons.
We determine the strength of the $U_A(1)$ anomaly term, which is found to 
give a significant portion of the diquark masses.
\end{abstract}
%
%\pacs{ }
%
\maketitle
%
% @@ =======================================================================
%
\section  {Introduction}
\label{sec:Introduction}

Recent developments of hadron spectroscopy has brought a completely new picture of 
hadrons. In particular, many unconventional hadron resonances have been found 
and become candidates of multiquark exotic states, which contain more than three 
quarks~\cite{Chen:2016qju,Hosaka:2016pey,Olsen:2017bmm,Liu:2019zoy}. 
X(3872) is a representative of such states.
While it is found in the charmonium ($c\bar c$) spectrum, 
it may not simply be a $c\bar c$, but is dominantly a tetraquark state, or a
$D\bar D^*$ molecular resonance. 
The other unconventional new resonances 
include charged hidden-charm mesons and hidden-charm pentaquark baryons.
It is urgent and important to reveal the composition and dynamics of these states.

In understanding their structures, we need a new dynamics that gives a strong correlation
among multiple quarks. 
Various types of subsystems, or clusters, have been proposed to form exotic multiquark states.
A candidate is color-singlet hadrons, which form a bound or resonance states.
Color-non-singlet clusters are more exotic and interesting. 
The simplest one, except for a constituent quark, is a diquark~\cite{Ida:1966ev,Lichtenberg:1967zz,Anselmino:1992vg}.
Various models with diquarks as ingredients have been proposed for explaining
masses and structures of multiquark hadrons~\cite{Jaffe:2003sg,Jaffe:2004ph,Hong:2004xn}.
Among many possible quantum numbers, it is known that 
the scalar ($0^+$) diquark with color $\bar 3$ and flavor $\bar 3$ is strongly
favored by quantum chromodynamics (QCD)~\cite{Hess:1998sd,Alexandrou:2006cq,Babich:2007ah,DeGrand:2007vu,Bi:2015ifa}, while the axial-vector ($1^+$) diquark with color $\bar 3$
and flavor $6$ also appear frequently in hadrons. Color $6$ diquarks are interesting also, but they may appear only in the multiquark states.

We consider a diquark effective theory for light quarks $q=(u, d, s)$. It is important 
to formulate their dynamics based on the chiral symmetry and its spontaneous 
breaking in QCD. Chiral symmetry plays key roles in understanding the low-lying
spectrum of mesons and baryons. In particular, the pion (and the other ground-state
pseudoscalar mesons) and its properties have revealed how chiral symmetry is 
broken in the QCD vacuum. The symmetry constrains the low-energy effective theory
of the pion very strongly.
In contrast, diquarks show new features and dynamics under the chiral symmetry~\cite{Ebert:1995fp,Nagata:2004ky,Hong:2004xn,Hong:2004ux,Nagata:2007di}.
Effective theories of diquarks were also explored in the context of color superconductivity at
high density QCD~\cite{Alford:1997zt,Rapp:1997zu}, where it is shown that the axial $U_A(1)$ anomaly plays an important role~\cite{Hatsuda:2006ps,Yamamoto:2007ah}.

Chiral symmetry of diquarks is closely related to the chiral representations of heavy-quark ($Q$) baryons ($Qqq$), such as singly charmed ($cqq$) or bottomed ($bqq$) baryons.
In fact, the roles of diquarks are
most prominently seen in the singly heavy baryon system~\cite{Kim:2011ut,Yoshida:2015tia,Jido:2016yuv,Kumakawa:2017ffl}. 
Chiral aspects of heavy baryon masses and decays were studied  in~\cite{Kawakami:2018olq,Kawakami:2019hpp}.

In this paper, we propose a chiral effective theory of scalar and pseudoscalar diquarks,
based on linear representations of $SU(3)_R\times SU(3)_L$ symmetry. 
Such an effective theory may be applied, once the parameters are determined by the known experimental data, to not only the heavy baryon systems, but also to tetraquarks,
and the other multiquark hadrons.
We write down a general
form of the effective Lagrangian for the scalar diquarks in the form of the linear sigma model.
Then the masses of the diquarks are identified at the tree level from the quadratic terms of the Lagrangian. 

It is shown that the leading  meson-diquark-diquark coupling term breaks axial $U_A(1)$ symmetry, while it keeps the chiral symmetry invariant. 
Such a term is supposed to come from $U_A(1)$ anomaly in QCD. 
Namely the flavor singlet axial-vector current for light quarks has a nonzero divergence due to 
quantum anomaly effect. It is known that this effect is connected to the coupling of light quarks to a nontrivial topological configuration of gluon, {\it i.e.,} instanton.
The $U_A(1)$ anomaly term with spontaneous chiral-symmetry breaking, {\it i.e.,} quark condensates, 
generates a diquark mass term, which behaves differently from the other mass terms.
We discuss how we can identify and determine the parameters of such a coupling term of
the diquark effective Lagrangian.

This paper is organized as follows.
In Sec.~II, we introduce diquarks and their local operator representation, and formulate chiral
effective theory in the chiral-symmetry limit.
In Sec.~III, explicit chiral-symmetry breaking due to the quark masses is introduced and its 
consequences are discussed.
In Sec.~IV, a numerical estimate is given for the parameters of the effective theory. 
We use the diquark masses calculated in lattice QCD and also the experimental values 
of the singly heavy baryons.
In Sec.~V, a conclusion is given.

\section{Diquark Effective Theory}

\subsection{Diquark operators in flavor $\bm{SU(3)}$ symmetry}

In order to study the transformation properties of the diquark systems, we consider properties of QCD composite operators made of two quark fields.
We employ the flavor $SU(3)$ basis for the quark operators, 
$q^a_{\alpha, i}$, where $a$ is color, $\alpha$ is Dirac, $i$ is flavor index of the quark.
Then local diquark operators are defined by
\begin{eqnarray}
&& (q^T_i \Gamma q_j)=  (q^{a}_{\alpha,i} \,(\Gamma)_{\alpha\beta}\, q^b_{\beta,j}),
\end{eqnarray}
where $T$ denotes the transpose for the Dirac index and $\Gamma$ is
a relevant combination of the Dirac gamma matrices. 
Possible combinations satisfying the Lorentz covariance and the Pauli principle for quarks are 
given in Table \ref{tab:diquark-1}. 

\begin{table}[htb]
\begin{center}
\begin{tabular}{c|l|c|c|c|c}
&&$J^{\pi}$ & color& flavor& $^{2S+1}L_J$\\
\hline\hline
1　&$(q^T C q)^{\bar 3}_A$& $0^-$ & $\bar 3$& $\bar 3$& $^3P_0$\\
2　&$(q^T C\gamma^5 q)^{\bar 3}_A $ & $0^+$ & $\bar 3$& $\bar 3$ & $^1S_0$\\
3　&$(q^T C\gamma^{\mu}\gamma^5 q)^{\bar 3}_A $ & $1^-$ & $\bar 3$& $\bar 3$& $^3P_1$\\
\hline
4　&$(q^T C\gamma^{\mu} q)^{\bar 3}_S$ & $1^+$ & $\bar{3}$& $6$ & $^3S_1$\\
5　&$(q^T C\sigma^{\mu\nu} q)^{\bar 3}_S $ & $1^+, 1^-$ & $\bar{3}$& $6$ & $^3D_1, {}^1P_1$\\
\hline
6　&$(q^T C q)^{6}_S$& $0^-$ & $6$& $6$ & $^3P_0$\\
7　&$(q^T C\gamma^5 q)^{6}_S $ & $0^+$ & $6$& $6$& $^1S_0$\\
8　&$(q^T C\gamma^{\mu}\gamma^5 q)^{6}_S$ & $1^-$ & $6$& $6$& $^3P_1$\\
\hline
9　&$(q^T C\gamma^{\mu} q)^{6}_A $ & $1^+$ & $6$& $\bar 3$& $^3S_1$\\
10　&$(q^T C\sigma^{\mu\nu} q)^{6}_A $ & $1^+, 1^-$ & $6$& $\bar 3$& $^3D_1, {}^1P_1$\\
\hline\hline
\end{tabular}
\end{center}
\caption{Local diquark operators. $C=i\gamma^0\gamma^2$ is the charge conjugation Dirac matrix.
The rightmost column shows the relevant quantum states for two (nonrelativistic) quarks.}
\label{tab:diquark-1}
\end{table}%

The first five operators belong to the total color $\bar 3$ representation, {\it i.e.}, color antisymmetric combinations, while the lower five are in the color symmetric $6$ representation. The indices $S$ (symmetric) and $A$ (antisymmetric) designate the flavor symmetry.
In the last column, the corresponding nonrelativistic quark model assignments are given
of the spin and orbital angular momentum.

\subsection{Scalar and pseudo-scalar diquarks in chiral $\bm{SU(3)_R\times SU(3)_L}$ symmetry}

In this paper, we concentrate on the scalar and pseudoscalar diquarks from the viewpoint of chiral symmetry.
More specifically, we consider the first two states, Nos.~1 and 2, from Table \ref{tab:diquark-1}, which have
spin $0$, color $\bar 3$ and flavor $\bar 3$. We here see that these two diquarks are chiral partners to each other, {\it i.e.}, they belong to the same chiral representation and therefore they would be degenerate if the chiral symmetry is not broken.

To see this, using the chiral projection operators, 
$P_{R,L}\equiv (1\pm\gamma_5)/2$,
we define the right quark, $q^a_{R,i} =P_R \, q^a_{i }$, as a (3,1) representation of chiral $SU(3)_R\times SU(3)_L$ symmetry, and the left quark, $q^a_{L,i} =P_L \, q^a_{i }$, 
as a (1,3) representation.
Explicitly, they transform as
\begin{eqnarray}
%&& q^a_{R,i} =P_R \, q^a_{i } , \quad q^a_{L,i} =P_L \, q^a_{i } , \\
&& q^a_{R,i} \to (U_R)_{ij} q^a_{R,j},\quad U_{R}\in SU(3)_{R}\\
&& q^a_{L,i} \to (U_L)_{ij} q^a_{L,j}, \quad U_{L}\in SU(3)_{L}.
\end{eqnarray}
Then we construct ``right'' and ``left'' spin-0 diquark operators
with color $\bar 3$ as
\begin{eqnarray}
&& d^a_{R,i}\equiv \epsilon^{abc}\epsilon_{ijk} (q^{bT}_{R,j} \,C \, q_{R,k}^c),
\\
&& d^a_{L,i}\equiv \epsilon^{abc}\epsilon_{ijk} (q^{bT}_{L,j} \, C \,q_{L,k}^c).
\end{eqnarray}
It is straightforward to show that $d_R$ and $d_L$
belong to chiral $(\bar 3, 1)$ and $(1, \bar 3)$
representation, respectively, and transform as
\begin{eqnarray}
&& d^a_{R,i} \to d^a_{R,j}\,(U^{\dagger}_{R})_{ji}, \quad (\bar 3,1),\\ 
&&  d^a_{L,i} \to d^a_{L,j}\,(U^{\dagger}_{L})_{ji}, \quad (1,\bar 3). 
\end{eqnarray}
The other diquark operators are also decomposed in the chiral basis similarly, as shown in Table~\ref{tab:diquark-chiral}.
First two of them are the diquarks that we concern in this paper.
\begin{table}[htp]
\begin{center}
\begin{tabular}{lccc}
\hline\hline
&spin & color& chiral\\
\hline
$d_{R,  i}^{a}=\epsilon_{abc}\epsilon_{ijk}(q_{R,j}^{bT} C q_{R,k}^c)$& $0$ & $\bar 3$& $(\bar 3,1)$\\
$d_{L,  i}^{a}=\epsilon_{abc}\epsilon_{ijk}(q_{L,j}^{bT} C q_{L,k}^c)$ & $0$ & $\bar 3$& $(1,\bar 3)$\\
$d_{(i,j)}^{a,\mu}=\epsilon_{abc} (q_{L,i}^{bT} C\gamma^{\mu}q_{R,j}^c)$ & $1$ & $\bar 3$& $(3,3)$\\
$d_{R\{ij\}}^{a,\mu\nu}=\epsilon_{abc}(q_{R,i}^{bT} C\sigma^{\mu\nu} q_{R,j}^c)$ & $1$ & $\bar{3}$& $(6,1)$\\
$d_{L\{ij\}}^{a,\mu\nu}=\epsilon_{abc}(q_{L,i}^{bT} C\sigma^{\mu\nu} q_{L,j}^c)$ & $1$ & $\bar{3}$& $(1,6)$\\
\hline
$\tilde d_{R\{ij\}}^{\{ab\}}=(q^{aT}_{R,i} C q^b_{R,j}+ q^{aT}_{R,j} C q^b_{R,i})$& $0$ & $6$& $(6,1)$ \\
$\tilde d_{L\{ij\}}^{\{ab\}}=(q^{aT}_{L,i} C q^b_{L,j}+ q^{aT}_{L,j} C q^b_{L,i})$ & $0$ & $6$& $(1,6)$\\
$\tilde d_{(i,j)}^{\{ab\},\mu} = (q^{aT}_{L,i} C\gamma^{\mu}q^b_{R,j}+q^{bT}_{L,i} C\gamma^{\mu}q^a_{R,j})$ & $1$ & $6$& $(3,3)$\\
$\tilde d_{R,i}^{\{ab\},\mu\nu}= \epsilon_{ijk} (q^{aT}_{R,j} C\sigma^{\mu\nu} q^b_{R,k}) $ & $1$ & $6$& $(\bar 3,1)$\\
$\tilde d_{L,i}^{\{ab\},\mu\nu}= \epsilon_{ijk} (q^{aT}_{L,j} C\sigma^{\mu\nu} q^b_{L,k}) $ & $1$ & $6$& $(1,\bar 3)$\\
\hline\hline
\end{tabular}
\caption{Local diquark operators in the chiral basis.}
\end{center}
\label{tab:diquark-chiral}
\end{table}

%\begin{table}[htp]
%\begin{center}
%\begin{tabular}{l|c|c|c}
%&spin & color& chiral\\
%\hline\hline
%$d_{R,  i}^{  a}=\epsilon_{abc}\epsilon_{ijk}(q_{R,j}^{bT} C q_{R,k}^c)$& $0$ & $\bar 3$& $(\bar 3,1)$\\
%$d_{L,  i}^{  a}=\epsilon_{abc}\epsilon_{ijk}(q_{L,j}^{bT} C q_{L,k}^c)$ & $0$ & $\bar 3$& $(1,\bar 3)$\\
%$d_{(i,j)}^{  a,\mu}=\epsilon_{abc} (q_{R,i}^{bT} C\gamma^{\mu}q_{L,j}^c)$ & $1$ & $\bar 3$& $(3,3)$\\
%$d_{R\{ij\}}^{  a,\mu\nu}=\epsilon_{abc}(q_{R,i}^{bT} C\sigma^{\mu\nu} q_{R,j}^c)$ & $1$ & $\bar{3}$& $(6,1)$\\
%$d_{L\{ij\}}^{  a,\mu\nu}=\epsilon_{abc}(q_{L,i}^{bT} C\sigma^{\mu\nu} q_{L,j}^c)$ & $1$ & $\bar{3}$& $(1,6)$\\
%\hline
%$\tilde d_{R\{ij\}}^{\{ab\}}=(q^{aT}_{R,i} C q^b_{R,j}+ q^{aT}_{R,j} C q^b_{R,i})$& $0$ & $6$& $(6,1)$ \\
%$\tilde d_{L\{ij\}}^{\{ab\}}=(q^{aT}_{L,i} C q^b_{L,j}+ q^{aT}_{L,j} C q^b_{L,i})$ & $0$ & $6$& $(1,6)$\\
%$\tilde d_{(i,j)}^{\{ab\}} = (q^{aT}_{R,i} C\gamma^{\mu}q^b_{L,j}+q^{bT}_{R,i} C\gamma^{\mu}q^a_{L,j})$ & $1$ & $6$& $(3,3)$\\
%$\tilde d_{R,i}^{\{ab\}}= \epsilon_{ijk} (q^{aT}_{R,j} C\sigma^{\mu\nu} q^b_{R,k}) $ & $1$ & $6$& $(\bar 3,1)$\\
%$\tilde d_{L,i}^{\{ab\}}= \epsilon_{ijk} (q^{aT}_{L,j} C\sigma^{\mu\nu} q^b_{L,k}) $ & $1$ & $6$& $(1,\bar 3)$\\
%\hline\hline
%\end{tabular}
%\end{center}
%\caption{Local diquark operators in the chiral basis.}
%\label{tab:diquark-chiral}
%\end{table}%

The diquarks given above are not eigenstates of parity. 
Using the parity transform of the quark operators,
${\cal P}: q^a_{iR}(t,\bm x) \to \gamma^0 q^a_{iL} (t,-\bm x)$,  $q^a_{iL}(t,\bm x) \to \gamma^0 q^a_{iR}(t,-\bm x)$,
the spin-0 diquarks are found to transform as
\begin{eqnarray}
%&& \hbox{parity}\quad  q^a_{iR} \to \gamma^0 q^a_{iL},  \quad q^a_{iL} \to \gamma^0 q^a_{iR}\\
&& {\cal P}: d^a_{R,i} \to -d^a_{L,i},  \quad d^a_{L,i} \to -d^a_{R,i}.
\end{eqnarray}
Thus the Lorentz scalar, $S$ ($J^\pi= 0^+$),  and pseudoscalar, $P$ ($0^-$), operators can be identified by
\begin{eqnarray}
&& S^a_{i} = \frac{1}{\sqrt{2}}(d^a_{R,i} - d^a_{L,i}) =  \frac{1}{\sqrt{2}} \epsilon^{abc}\epsilon_{ijk} (q^{bT}_{j} \,C \gamma_5\, q^c_{k}),
\label{Sdiquark}\\
&& P^a_{i} = \frac{1}{\sqrt{2}}(d^a_{R,i} + d^a_{L,i}) =  \frac{1}{\sqrt{2}}\epsilon^{abc}\epsilon_{ijk} (q^{bT}_{j} \,C \, q^c_{k}). \label{Pdiquark}
\end{eqnarray}
These relations show that the scalar and pseudoscalar diquarks, given as Nos.~1 and 2 of Table \ref{tab:diquark-1}, belong to $(\bar 3, 1)$ and $(1,\bar 3)$ representations of chiral symmetry.
Thus, we conclude that the scalar and pseudoscalar diquarks are chiral partners.

\subsection{Chiral Lagrangian in the chiral limit}

We now introduce the chiral ($\bar 3$,3) meson fields $\Sigma$, which contain nonet pseudoscalar and nonet scalar mesons. Their chiral transform is given by
\begin{eqnarray}
&& \Sigma_{ij} \equiv \sigma_{ij} +i\pi_{ij} \to U_{L,ik} \Sigma_{km} U_{R,mj}^\dagger \quad (\bar 3, 3)
\end{eqnarray}
where $\sigma$ represents a scalar nonet, and $\pi$ a pseudoscalar nonet. $\Sigma$ transforms under the spatial inversion as ${\cal P}: \Sigma \to \Sigma^\dagger$.
Chiral symmetry is spontaneously broken by the vacuum state, which is represented by the 
vacuum expectation value (VEV) of the scalar field $\sigma$ as
\begin{eqnarray}
&& \langle\Sigma_{ij}\rangle=\langle\sigma_{ij}\rangle=f\delta_{ij} ,\quad \langle\pi_{ij}\rangle=0
\label{SSB}
\end{eqnarray}
where $f$ is the pion decay constant. 
For this vacuum, $\pi_{ij}$ is the nonet of massless Nambu-Goldstone bosons.
To read off the conventional meson contents from $\Sigma$, 
one uses the Gell-Mann matrices, $\lambda_p$, as
\begin{eqnarray}
%&& \Sigma_{ij} \equiv \frac{(\lambda_p)_{ij}}{2} (\sigma_p+i\pi_p)\\
%&& \sigma_p = \frac{1}{2}{\rm Tr}\left[ \lambda_p(\Sigma +\Sigma^\dagger)\right], \\
%&& \pi_p = \frac{1}{2i}{\rm Tr}\left[ \lambda_p(\Sigma -\Sigma^\dagger)\right] .
&& \Sigma_{ij} \equiv (\lambda_p)_{ij} (\sigma_p+i\pi_p)\\
&& \sigma_p = \frac{1}{4}{\rm Tr}\left[ \lambda_p(\Sigma +\Sigma^\dagger)\right], \\
&& \pi_p = \frac{1}{4i}{\rm Tr}\left[ \lambda_p(\Sigma -\Sigma^\dagger)\right] .
\end{eqnarray}
Note that $\sigma$ and $\pi$ contain the flavor singlet components, 
for which we use $\lambda_0=\sqrt{2/3} {\bm 1}$ with the unit matrix ${\bm 1}$.

We are ready to present an effective Lagrangian in the chiral limit as
\begin{eqnarray}
&& {\cal L}= \Dc_{\mu} d_{R,i} \, (\Dc^{\mu} d_{R,i})^\dagger + {\cal D}_{\mu} d_{L,i} \,({\cal D}^{\mu} d_{L,i})^\dagger \nonumber\\
&&\quad - m_{0}^2 (d_{R,i} d_{R,i}^\dagger  +d_{L,i} d_{L,i}^\dagger) \nonumber\\
&& \quad -\frac{m_{1}^2}{f} (d_{R,i}\Sigma_{ij}^\dagger d_{L,j}^\dagger +d_{L,i}\Sigma_{ij} d_{R,j}^\dagger)\nonumber\\
&&\quad - \frac{m_{2}^2}{2f^2} \epsilon_{ijk}\epsilon_{\ell mn} (d_{R,k} \Sigma_{\ell i}\Sigma_{mj} d_{L,n}^\dagger 
 + d_{L,k} \Sigma^\dagger_{\ell i}\Sigma^\dagger_{mj} d_{R,n}^\dagger)\nonumber\\
%&&\quad + \frac{m_3^2}{f^2} \left[ d_{R,i}
% (\Sigma_{ik}^\dagger\Sigma_{kj}-\delta_{ij} {\rm Tr}[\Sigma^{\dagger}\Sigma]) d_{R,j}^\dagger \right.\nonumber\\
%&& \qquad \left. +d_{L,i}(\Sigma_{ik}\Sigma_{kj}^\dagger -\delta_{ij} {\rm Tr}[\Sigma\Sigma^{\dagger}]) d_{L,j}^\dagger \right]\nonumber\\
&& \quad + \frac{1}{4}{\rm Tr}\left[\partial^{\mu}\Sigma^{\dagger} \partial_{\mu}\Sigma\right] + V(\Sigma).
\label{Lagrangian}
\end{eqnarray}
Here we truncate the interaction terms with more than two $\Sigma$'s.
We also omit $\Sigma\Sigma^\dagger$ and $\Sigma^{\dagger}\Sigma$ terms,
since they do not contribute to the mass difference between the $0^+$ and $0^-$ states.
$V(\Sigma)$, which is not shown explicitly, denotes the interaction potential terms 
for the meson fields that cause the spontaneous symmetry breaking, Eq.~(\ref{SSB}). 
Hereafter, we will omit the kinetic and potential parts of the mesons, because we consider only
the mean fields of the mesons. 

As the diquark is not a color-singlet state, 
we have introduced a color-gauge-covariant derivative in Eq.~(\ref{Lagrangian}), 
$\Dc_{\mu}= \partial_{\mu} +ig T^{\alpha} G^{\alpha}_{\mu}$, 
with $G_{\mu}$ being the gluon field, and 
$T^{\alpha}$ the color $SU(3)$ generator for the $\bar 3$ representation.
All the color indices are contracted and not explicitly written. 
The kinetic energy term of the gluons fields is also omitted.

It is easy to check that the Lagrangian, Eq.(\ref{Lagrangian}), is chiral invariant and parity conserving. We may rewrite the Lagrangian in terms of the parity eigenstates, Eqs.~(\ref{Sdiquark}) and (\ref{Pdiquark}), as
\begin{eqnarray}
&& {\cal L}= \Dc_{\mu} S_i \, (\Dc^{\mu} S_i)^\dagger + {\cal D}_{\mu} P_i \,({\cal D}^{\mu} P_i)^\dagger 
\nonumber\\
&& \quad - m_{0}^2 (S_i S_i^\dagger  + P_i P_i^\dagger) \nonumber\\
&&\quad -\frac{m_{1}^2}{f} (-S_i \sigma_{ij} S_j^\dagger + P_i \sigma_{ij} P_j^\dagger
- i S_i \pi_{ij} P_j^\dagger+ i P_i \pi_{ij} S_j^\dagger ) \nonumber\\
&& \quad
- \frac{m_{2}^2}{2f^2} \epsilon_{ijk}\epsilon_{\ell mn} \left[
-S_{k}(\sigma_{\ell i}\sigma_{mj} - \pi_{\ell i}\pi_{ mj})S_{n}^\dagger \right.\nonumber\\
&&\quad +P_{k}(\sigma_{\ell i}\sigma_{mj} - \pi_{\ell i}\pi_{ mj})P_{n}^\dagger 
+ i S_{k}(\pi_{\ell i}\sigma_{mj} + \sigma_{\ell i}\pi_{ mj})P_{n}^\dagger\nonumber\\
&& \quad\left.  
- i P_{k}(\pi_{\ell i}\sigma_{mj} + \sigma_{\ell i}\pi_{ mj})S_{n}^\dagger\right] 
\label{Lagrangian-Parity}
\end{eqnarray}

\subsection{Masses of the diquarks and generalized Goldberger-Treiman relation}

In the mean field approximation, keeping the $SU(3)$ symmetry, 
$\langle\Sigma_{ij}\rangle=f \delta_{ij}$, 
the masses of the diquarks are read from 
\begin{eqnarray}
&& {\cal L}_{\rm mass}= - m_{0}^2 (d_{R,i} d_{R,i}^\dagger  +d_{L,i} d_{L,i}^\dagger)\nonumber\\
&& \quad-(m_{1}^2+m_{2}^2) (d_{R,i} d_{L,i}^\dagger +d_{L,i} d_{R,i}^\dagger),
\label{mass terms}
\end{eqnarray}
which leads to the mass matrix for $(d_{R,i}, d_{L,i})$ as
\begin{eqnarray}
&& M^2= \begin{pmatrix} {m_{0}^2& m_{1}^2+m_{2}^2 \cr m_{1}^2+m_{2}^2 &m_{0}^2} \end{pmatrix}
\end{eqnarray}
Diagonalizing the mass matrix, we obtain the mass eigenstates,
\begin{eqnarray}
&& S^{a}_{i} = \frac{1}{\sqrt{2}}(d^{a}_{R,i} - d^{a}_{L,i}) \nonumber\\
&&\qquad \longrightarrow M(0^+)=\sqrt{m_{0}^2 - m_{1}^2-m_{2}^2}, \label{diquark mass +}\\
&&  P^{a}_{i} = \frac{1}{\sqrt{2}}(d^{a}_{R,i} + d^{a}_{L,i}) \nonumber\\
&&\qquad\longrightarrow M(0^-)=\sqrt{m_{0}^2 + m_{1}^2+m_{2}^2},\label{diquark mass -}
\end{eqnarray} 
which are also the eigenstates of the parity.

Now it is clear how the diquark masses are generated from the spontaneous chiral-symmetry breaking (SCSB). 
In the regime of complete chiral restoration, the $0^+$ and $0^-$ diquarks are degenerate with the mass $m_0$. This is the limit where all the hadrons are subjected to belong to a parity doublet.
In the ordinary vacuum of QCD, the SCSB resolves their degeneracies.
In the present case, the mass splitting is given by the $m_1^2$ and $m_2^2$ terms.

It should be noted that the diquarks are bosons and their chiral behaviors are different from fermions. In the linear sigma model for baryons, if we assign the chirality of the baryon according to the chirality of quarks (naive choice), then the baryon mass should vanish in the chiral-symmetric limit. 
In this case, the baryon mass comes only from SCSB.
It was, however, shown that the mirror assignment of chirality ($L\leftrightarrow R$ reversed) of the baryon is possible, and then the chiral symmetric mass term is allowed. Realistic baryons may be a mixing of these two assignments~\cite{Detar:1988kn,Jido:1998av,Jido:1999hd}. 

In contrast, the chiral representation of the spin-0 diquarks allows both the chiral symmetric and SCSB mass terms simultaneously. Namely, the $m_0^2$ term is independent from SCSB, 
while the $m_1^2$ and $m_2^2$ terms contribute to the baryon masses only when the chiral symmetry is spontaneously broken.

%\subsection{Generalized Goldberger-Treiman relation}

As is shown in Eq.(\ref{Lagrangian-Parity}), the $m_1^2$ and $m_2^2$ terms of the Lagrangian 
describe the meson-diquark interactions.
The $\pi-S-P$ vertex terms are given by
\begin{eqnarray}
&& {\cal L}_{\pi SP} = \frac{i m_{1}^2}{f} ( S_i \pi_{ij} P_j^\dagger- P_i \pi_{ij} S_j^\dagger ) \nonumber\\
&&\quad -\frac{im_2^2}{f^2}\epsilon_{ijk}\epsilon_{\ell mn} 
(S_{k}\pi_{\ell i}\langle\sigma_{mj}\rangle P_{n}^\dagger 
-P_{k}\pi_{\ell i}\langle\sigma_{mj}\rangle S_{n}^\dagger)\nonumber\\
&& =\frac{i(m_1^2+m_2^2)}{f} (S_i \pi_{ij} P_j^\dagger-P_i \pi_{ij} S_j^\dagger)\nonumber\\
&&\quad -\frac{im_2^2}{f} {\rm Tr} [\pi] (S_i P_i^\dagger-P_i S_i^\dagger)\nonumber\\
&& = \frac{i(m_1^2+m_2^2)}{f} \pi_p(S \lambda_p P^\dagger-P \lambda_p S^\dagger)\nonumber\\
&&\quad -\frac{3im_2^2}{f} \pi_{0} (S \lambda_0 P^\dagger-P \lambda_0  S^\dagger)
\end{eqnarray}
where $\pi_{0}=\eta_1=\sqrt{6} {\rm Tr} [\pi]$ is the singlet pseudoscalar meson.

We then obtain the relation between the octet-meson-diquark couplings and the mass differences of diquarks, as a generalized Goldberger-Treiman (GT) relation,
\begin{eqnarray}
&& g_{\pi SP} \equiv \frac{m_1^2+m_2^2}{f}= \frac{M^2(0^-)-M^2(0^+)}{2f}. 
\end{eqnarray} 
This relation can also be derived from the conservation of the flavor-octet axial-vector currents and the existence of the massless Nambu-Goldstone bosons, $\pi_p$ ($p=1,\ldots 8$).
Note that this coupling is a nonderivative, $S$-wave, coupling that describes the mesonic decay of the negative-parity excited heavy baryon into the positive-parity ground states. 
%In general, $S$-wave decay of a negative-parity baryon has a large decay width.

On the other hand, the coupling constant of the singlet (eta) meson $\pi_0=\eta_1$ is given by
\begin{eqnarray}
&& g_{\pi_0 SP} = \frac{m_1^2-2m_2^2}{f}. 
\end{eqnarray}
This relation is not a GT relation and is not directly derived from the symmetry because the axial $U_A(1)$ is explicitly broken and the singlet 
$\eta_1$ is not a massless Nambu-Goldstone boson as is discussed in the next section.

\subsection{$\bm{U_A(1)}$ anomaly}

So far, we have considered the chiral $SU(3)_R\times SU(3)_L$ symmetry. 
The given Lagrangian, Eq.(\ref{Lagrangian}), is invariant under the $SU(3)_R\times SU(3)_L$ transform.
However, the QCD Lagrangian for massless quarks has another axial symmetry, $U_A(1)$ symmetry, which counts 
the difference of right and left quarks, regardless of flavor.
In fact, this symmetry is not realized in the hadron spectrum due to anomaly.
It is known that the instanton, a topologically nontrivial configuration of the gluon field in the Euclidean 4-dimensional space-time, plays a leading role in the $U_A(1)$ breaking. There the light quarks, ($u,d,s$) couples to the instanton in an axial-symmetry breaking manner~\cite{tHooft:1986ooh,tHooft:1976snw}.

In low-energy effective field theories, the $U_A(1)$ anomaly can be taken into account as an effective 
symmetry-breaking term. For instance, for the light meson sector, it is given as an extra
term like $g_D\, {\rm det} \,(\Sigma+\Sigma^\dagger)$ (Kobayashi-Maskawa-'t~Hooft (KMT) term)~\footnote{We suppose that this term is included in the meson potential term in Eq.~(\ref{Lagrangian}).}.
It makes the flavor singlet $\eta_1(=\pi_0)$ massive, while the octet $\eta_8(=\pi_8)$ is massless in the chiral limit~\cite{Hatsuda:1994pi,Kuroda:2019jzm,Kono:2019aed}.

We similarly consider
the $U_A(1)$ anomaly for the diquark effective theory.
It happens that the $m_1^2$ term of Eq.(\ref{Lagrangian}),
\begin{eqnarray}
&& {\cal L}_{m1}= - \frac{m_{1}^2}{f} (d_{R,i}\Sigma_{ij}^\dagger d_{L,j}^\dagger +
d_{L,i}\Sigma_{ij} d_{R,i}^\dagger)
\label{SUA1}
\end{eqnarray}
breaks $U_A(1)$ symmetry~\cite{Hatsuda:2006ps,Yamamoto:2007ah}.
It is easy to prove that each term contains three left and 
three right quarks with antisymmetric flavor indices.
Thus the term is proportional to ${\rm det}_{i,j} (q_{R,j}q^\dagger_{L,i} + q_{L,j}q^\dagger_{R,i} )$.
This is nothing but the %Kobayashi-Maskawa-'t~Hooft (KMT) 
determinant interaction, which is known to come 
from the instanton-light-quark couplings and breaks the $U_A(1)$ symmetry.
In fact, by using the Fierz transformation, one can explicitly show 
\begin{eqnarray}
&&d_{R,i}\Sigma_{ij}^\dagger d_{L,j}^\dagger \propto d^b_{R,i} (q^a_{R,i} \bar q^a_{L,j}) d^{b\dagger}_{L,j}  \nonumber\\
&& \quad= \epsilon_{bcd}\epsilon_{ipq} (q^{cT}_{R,p} \,C \, q^d_{R,q})
\epsilon_{bef}\epsilon_{jrs} (q^{eT}_{L,r} \, C \,q^f_{L,s})^\dagger
 (q^a_{R,i} \bar q^a_{L,j})\nonumber\\
 &&\quad  =
12\, {\rm det}_{ij}(\bar q^a_{L,i} q^a_{R,j})
%&&(\bar q_{1}q_{2})(\bar q_3 q_4) =
%\frac{1}{4} \left( (\bar q_{1}C \bar q_{3})(q_4 C q_2)+ (\bar q_{1}\gamma_5 C \bar q_{3})
%(q_4 C\gamma_5 q_2) \right)\\
%&& (\bar q_{iR}q_{\ell L})(\bar q_{jR}q_{mL}) =
%\frac{1}{2} (\bar q_{iR}C \bar q_{jR})(q_{mL} C q_{\ell L})\\
%\nonumber\\
%&& {\rm det}(\bar q_{iR} q_{jL}) = \frac{1}{12} (d^a_{kR})^{\dagger} d^a_{nL} (\bar q^c_{kR} q^c_{nL})
%\propto \frac{1}{12} (d^a_{kR})^{\dagger} d^a_{nL} \Sigma_{nk} = \frac{1}{12} d_L \Sigma d_R^{\dagger}
\end{eqnarray} 
It is clear that this term breaks $U_A(1)$ symmetry as the numbers of left and right quarks in each term are different. On the other hand, it keeps $SU(3)_R\times SU(3)_L$ invariant, because the 
flavor determinant is invariant under the $SU(3)$ transform.

\begin{figure*}[htb]
\begin{center}
\includegraphics[bb=0 0 648 290, height=6cm]{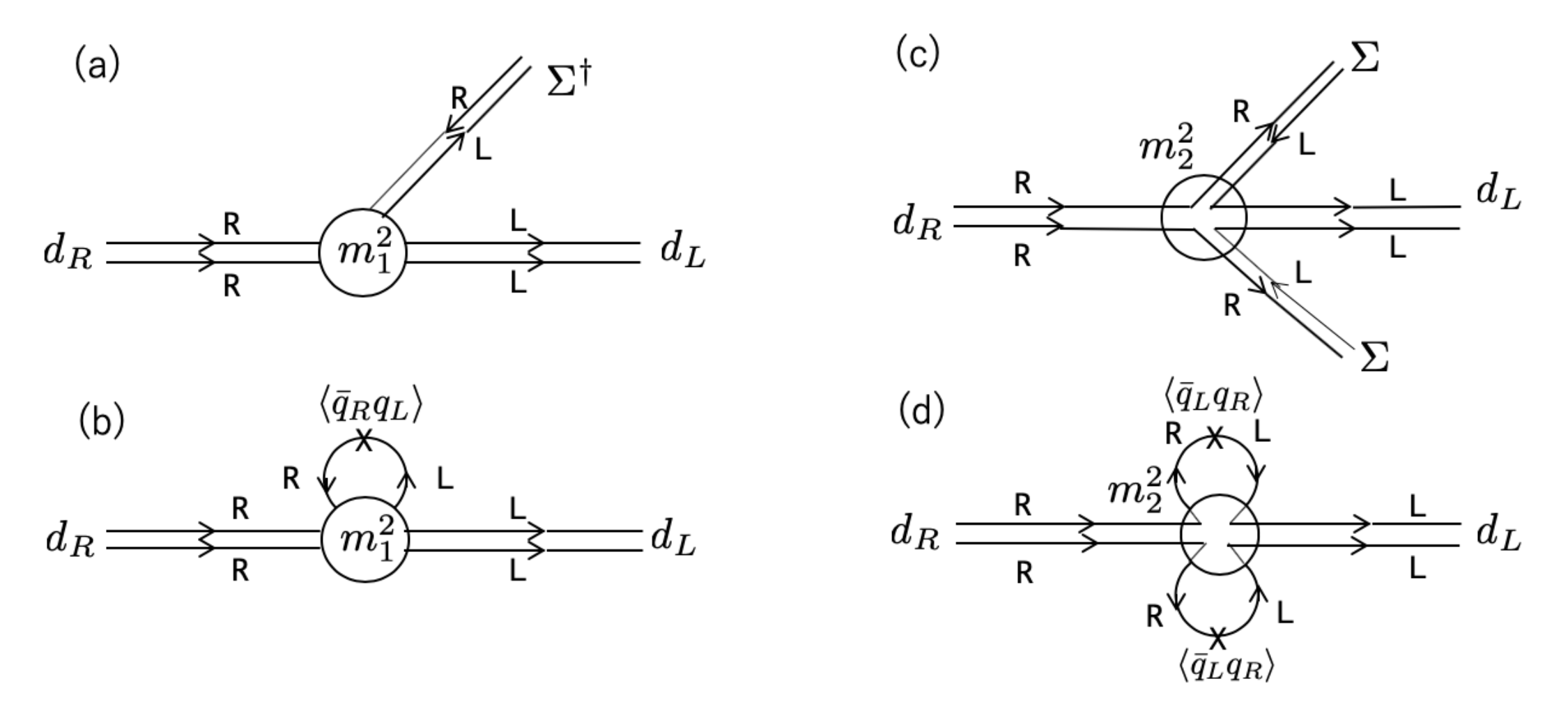}
\caption{Quark line representations of the diquark interaction terms. (a) $m_1^2$ term, (b) $m_1^2$ term with quark condensate, (c) $m_2^2$ term, and (d) $m_2^2$ term with quark condensate.}
\label{diquark}
\end{center}
\end{figure*}

Figure 1 shows the chiral properties of the vertices of the Lagrangian, Eq.~(\ref{Lagrangian}). The $m_1^2$ term contains a six-quark vertex induced by the instanton, while the $m_2^2$ term does not break the $U_A(1)$ symmetry.
The $U_A(1)$ anomaly effects will arise to the diquark mass and interaction only when
the chiral symmetry is spontaneously broken.

\subsection{Masses of the singly heavy baryons}

As the diquarks are not directly observed, 
we may instead consider a singly heavy baryon, 
a bound state of a diquark with a heavy quark $Q$ (charm or bottom)~\cite{Kawakami:2018olq,Kawakami:2019hpp}.
Corresponding baryons are $\Lambda_Q$, and $\Xi_Q$ with spin-parity $1/2^+$ and $1/2^-$.
We write down the effective Lagrangian for the $Qqq$ baryons 
with the one-to-one correspondence to Eq.(\ref{Lagrangian}), as
\begin{eqnarray}
&& {\cal L}_{Qqq\,\,\rm baryons} =
\bar S_{R,i} (iv\cdot \partial) S_{R,i} + \bar S_{L,i} (iv\cdot \partial) S_{L,i}\nonumber\\
&&  \quad - M_{B0} \left( \bar{S}_{R,i} \, S_{R,i}  + \bar{S}_{L,i} \, S_{L,i} \right)  \nonumber\\
 && \quad - \frac {M_{B1}}{f} (\bar S_{R,i}\Sigma^{T}_{ij} S_{L,j} +\bar S_{L,i}\Sigma^{T\dagger}_{ij} S_{R,j})
 \nonumber\\
&&\quad- \frac{M_{B2}}{2f^2} \epsilon_{ijk}\epsilon_{\ell mn} (\bar S_{L,k}\Sigma^T_{\ell i}\Sigma^T_{mj} S_{R,n}
 + \hbox{h.c.}),
\end{eqnarray}
where $S_{L/R,i}$ denotes the effective field for the $Qqq$ baryon multiplets with the left/right $\bar 3$ representation carrying the velocity $v^\mu$.
%\cmh{
%and $v^\mu$ the velocity of the heavy baryons.
We note that these $S_{L/R,i}$ fields are the heavy hadron effective fields for the fluctuation modes satisfying $v^\mu \gamma_\mu \, S_{L/R,i} = S_{L/R,i}$.%
\footnote{
The fields $S_{L/R,i}$ are related to the heavy baryon Dirac field operators $B_{L/R,i}$ as $B_{L/R,i}  = \sum_{v^\mu} e^{- i M_Q v^\mu x_\mu } B_{L/R,i (v)} $ and $ S_{L/R,i} = P_{+} B_{L/R,i (v)}$, where $P_+$ is the projection operator defined as $P_+ = ( 1 + v^\mu \gamma_\mu )/2$ and $M_Q$ is the heavy-quark mass.
} 
We may redefine the fields to eliminate the $M_{B0}$ term as $S_{L/R,i} \to e^{- i M_{B0} v^\mu x_\mu} S_{L/R,i}$, but we keep this term to see the explicit correspondence to the diquark Lagrangian in Eq.~(\ref{Lagrangian}).
%}
 The $M_{B1}$ term is the one that breaks the $U_A(1)$ symmetry. 
This term generates anomalous meson-baryon couplings,
the $S$-wave $\Xi_Q(1/2^-) \Xi_Q(1/2^+) \pi$ and $\Xi_Q(1/2^+) \Lambda_Q (1/2^-)  K$ couplings.
%\cmh{
The $M_{B2}$ term is invariant under $U_A(1)$ symmetry transformation in addition to the chiral $SU(3)_R \times SU(3)_L$ symmetry transformation.
%}

%\cmh{
As in Eq.~(\ref{SSB}), the VEV of $\Sigma$ field causes the spontaneous chiral-symmetry breaking.
Then, from the above Lagrangian, the masses of the baryons with positive and negative parities are given by
\begin{eqnarray}
M_{B+} &=&  M_Q+ M_{B0} - M_{B1} - M_{B2} \ , \nonumber\\
M_{B-} &=&  M_Q+ M_{B0} + M_{B1} + M_{B2} \ . \label{heavy baryon masses}
\end{eqnarray}
where $M_Q$ is the heavy-quark mass.
As stated above, one can absorb $M_{B0}$ into the redefinition of $M_Q$, which implies that it is impossible to distinguish $M_{B0}$ and $M_Q$.

Now, let us assume that the binding energies between the heavy quark $Q$ and the diquark $qq$ are the same for the baryons with positive and negative parities.
Then, the mass difference between two baryons is determined by the mass difference of the relevant diquarks. 
By comparing the formulas in Eq.~(\ref{heavy baryon masses}) with the ones for diquark masses in Eqs.~(\ref{diquark mass +}) and (\ref{diquark mass -}), the mass difference of chiral partners of the singly heavy baryons is related to the diquark mass parameter as
\begin{eqnarray}
&& M_{B-} - M_{B+}  =  M(0^-) - M(0^+) \nonumber\\
&& \quad = \sqrt{ m_0^2 + m_1^2 + m_2^2 } - \sqrt{ m_0^2 - m_1^2 - m_2^2 } \ .
\end{eqnarray}
%}

\section{Quark masses, $\bm{SU(3)}$ breaking}

It is important to include the effects of the explicit breaking of chiral symmetry 
due to the quark masses, $(m_u, m_d, m_s)$, which are not zero nor equal.
The mass hierarchies of the light mesons and baryons reflect the $SU(3)$ breaking 
due to the quark masses. In effective theories, this effect comes either in the choices
of the parameters, or with extra terms with explicit breaking or both.

\subsection{Chiral Lagrangian with explicit symmetry breaking}

In the linear sigma model, we consider the effective quark mass generated by 
the current quark mass and the spontaneous chiral-symmetry breaking.
Namely, with the condensation 
of $\sigma$, the quarks acquire an effective mass of $\sim 300 - 500$ MeV.
\begin{eqnarray}
&& -m_i \bar q_i q_i \longrightarrow -(m_i +g_s \langle \sigma_{ii} \rangle) \bar q_i q_i
\end{eqnarray}
where $m_i$ is the current quark mass of the $i$th flavor and $g_s(\sim 3)$ denotes the coupling of the scalar meson, $\sigma$, to the quark.
We choose $\langle\sigma_{11}\rangle=\langle\sigma_{22}\rangle =f_{\pi}\sim 92$ MeV and $\langle\sigma_{33}\rangle =f_{s}=2f_K-f_{\pi}\sim 128$ MeV.

In a more general form, using the quark mass matrix,
${\cal M}\equiv {\rm diag} (m_u, m_d, m_s)$,
and the VEV of $\Sigma$, we write the effective mass as
\begin{eqnarray}
&& {\cal M}_{\rm eff}  = {\cal M} + g_s \langle \Sigma \rangle \simeq  (g_s f_{\pi}) \,\,{\rm diag} \{ 1,1, A\}, \\
&& A \equiv \frac{f_{s}}{f_{\pi}} \left(1+ \frac{m_s}{g_s f_{s}}\right) >1 .
\label{Apara}
\end{eqnarray}
Here we neglect $u$ and $d$ quark masses $\sim$2 and 5 MeV, respectively,
as they are much smaller than $g_s f_\pi \sim 300$ MeV, while for $m_s\sim100-200$ MeV, 
$A \sim 5/3$ gives a significant correction.

Now we consider the symmetry breaking in the interaction terms of the Lagrangian, Eq.(\ref{Lagrangian}).
The above consideration leads us to a prescription that the explicit symmetry breaking is introduced by the replacement,
\begin{eqnarray}
&& \Sigma \longrightarrow  \tilde\Sigma \equiv \Sigma + {\cal M}/g_s.
\end{eqnarray}
This is justified because every mass insertion to a quark line in Feynman diagrams can 
have chiral-symmetry breaking $\langle\bar q q\rangle$ condensate in the same line.

Then this prescription gives the Lagrangian that includes explicit chiral-symmetry breaking as
\begin{eqnarray}
&& {\cal L}_{\rm int}= 
-\frac{m_{1}^2}{f_{\pi}} (d_{R,i}\tilde\Sigma_{ij}^\dagger d_{L,j}^\dagger +d_{L,i}\tilde\Sigma_{ij} d_{R,j}^\dagger)\nonumber\\
&&\quad - \frac{m_{2}^2}{2f_{\pi}^2} \epsilon_{ijk}\epsilon_{\ell mn} (d_{R,k} \tilde\Sigma_{\ell i} \tilde\Sigma_{mj} d_{L,n}^\dagger 
+ \hbox{h.c.}),
\label{SBLagrangian}
\end{eqnarray}
with $\displaystyle\tilde\Sigma_{ij}\equiv \Sigma_{ij} + \frac{1}{g_s}{\cal M}_{ij}$.

\subsection{Diquark masses with $\bm{SU(3)}$ breaking}

In the chiral-symmetry breaking vacuum, 
by replacing $\tilde\Sigma$ with its expectation value,
$ \langle\tilde\Sigma\rangle = {\cal M}_{\rm eff}/g_s =f_{\pi} {\rm diag} \left\{ 1,1,A\right\}$, 
in Eq.(\ref{SBLagrangian}), we can read off the mass terms, as
\begin{eqnarray}
&& {\cal L}_{\rm mass}
= - m_{0}^2 (d_{R,i} d_{R,i}^\dagger  +d_{L,i} d_{L,i}^\dagger)\nonumber\\
&& \quad  - (m_{1}^2+A m_2^2 ) (d_{R,1} d_{L,1}^\dagger +d_{L,1} d_{R,1}^\dagger
\nonumber\\
&&\qquad \quad +d_{R,2} d_{L,2}^\dagger +d_{L,2} d_{R,2}^\dagger) \nonumber\\
&& \quad - (Am_{1}^2+m_2^2 ) ( d_{R,3} d_{L,3}^\dagger +d_{L,3} d_{R,3}^\dagger) .
\label{SBLagrangian-mass}
\end{eqnarray}
The mass eigenstates are obtained by diagonalizing the mass matrix for each flavor,
\begin{eqnarray}
&& (M^2)_{1,2}= \begin{pmatrix} {m_{0}^2& m_{1}^2+Am_{2}^2 \cr m_{1}^2+Am_{2}^2 &m_{0}^2} \end{pmatrix} ,\\
&& (M^2)_3= \begin{pmatrix} {m_{0}^2& Am_{1}^2+m_{2}^2 \cr Am_{1}^2+m_{2}^2 &m_{0}^2} .
\end{pmatrix}
\end{eqnarray}
The eigenstates coincide with the scalar ($S_i$, $0^+$) and 
pseudoscalar ($P_i$, $0^-$) diquarks again
and their masses are given by
\begin{eqnarray}
&& M_1(0^+) = M_2(0^+) = \sqrt{m_0^2 - m_{1}^2 - Am_2^2} ,\label{M1P}\\
&& M_3(0^+) =  \sqrt{m_0^2 - Am_{1}^2 - m_2^2}, \label{M3P}\\
&& M_1(0^-) = M_2(0^-) = \sqrt{m_0^2 +m_{1}^2 +Am_2^2}, \label{M1M}\\
&& M_3(0^-) = \sqrt{m_0^2 + Am_{1}^2 +m_2^2} .\label{M3M}
\end{eqnarray}

Now one sees nontrivial hierarchy structures of the diquark masses. 
From Eqs.~(\ref{M1P})-(\ref{M3M}), one obtains 
\begin{eqnarray}
&& \left[ M_{1,2}(0^+) \right]^2 - \left[ M_{3}(0^+) \right]^2 
= \left[ M_{3}(0^-) \right]^2 - \left[ M_{1,2}(0^-) \right]^2 \nonumber\\
&&\qquad= ( A - 1) ( m_1^2 - m_2^2 ).
\label{rel-diff}
\end{eqnarray}
Note that the $i=1$ ($ds$) and $i=2$ ($su$) diquarks are
the ones with the strange quark, while the $i=3$ ($ud$) diquark is nonstrange.

Suppose that the $(ud)$ scalar diquark is lighter than the $(ds)$ and $(su)$ diquarks. 
This is a natural assumption which can be confirmed from the spectrum of the singly heavy baryons, $M(\Xi_Q=Qsu,Qds) > M(\Lambda_Q=Qud)$.
As $M_1$ and $M_2$ correspond to the isodoublet diquarks, we need to have
a mass hierarchy, $M_1(0^+)=M_2(0^+)>M_3(0^+)$. 
Now from Eq.~(\ref{rel-diff}) and $A>1$, $m_1^2>m_2^2$ is required.
Then for the negative-parity diquarks, we will have $M_1(0^-) =M_2(0^-)< M_3(0^-)$.
This is an inverse hierarchy, because for the negative-parity diquarks, the strange ones ($i=1$, 2)
are lighter than the nonstrange ($i=3$) one.

\section{Numerical estimates}

In order to determine the parameters of the effective Lagrangian, we need to have a few inputs.
Ideally, the masses of the diquarks are useful. There are several attempts of computing the diquark masses and spectrum in lattice QCD~\cite{Hess:1998sd,Alexandrou:2006cq,Babich:2007ah,DeGrand:2007vu,Bi:2015ifa}. 
As the diquark is not a color-singlet state, we need
either fixing the gauge on the lattice and measure the diquark masses, or placing a heavy color source to compensate the color of the diquark and measure the mass (energy) differences of the different diquark states.
Both the methods give qualitatively consistent results, in particular for the mass difference between the scalar diquark ($0^+$) and the axial-vector diquark ($1^+$), which is about $150-200$ MeV.

The other possible inputs are the masses of singly heavy baryons.
The bound states of a spin-0 diquark and a charm quark form
charmed baryons, such as $\Lambda_c^+$ ($cud$, $1/2^{\pm}$) and 
$\Xi_c^{+,0}$ ($csu$ or $csd$, $1/2^{\pm}$)~\cite{Liu:2009jc,Briceno:2012wt,Namekawa:2013vu,Brown:2014ena,Bali:2015lka,Alexandrou:2017xwd,Can:2019wts}. 
Assuming that the charm quark is a spectator, we can estimate the mass differences among the diquarks from those of the baryons.

Here we present two methods of determining the parameters, Method I (from lattice QCD) and II (from heavy baryon masses), in the following.

\subsection{Method I}

First, we take the diquark masses from a recent lattice QCD calculation for the Landau gauge in full QCD~\cite{Bi:2015ifa},
\begin{eqnarray}
&& M_3(0^+) = 725 \mbox{ MeV,} \label{M3P val 2}\\
&& M_3(0^-) = 1265 \mbox{ MeV,} \label{M3M val 2}\\
&& M_{1,2}(0^+) = 906 \mbox{ MeV.}\label{M1P val 2}
\end{eqnarray}
Then from Eq.~(\ref{rel-diff}), we obtain
\begin{eqnarray}
&& M_{1,2}(0^-) = 1142\mbox{ MeV.}
\end{eqnarray}
By using the observed $\Lambda_c(1/2^+)$ mass given in Eq.~(\ref{Lambda c mass})
as an input, we estimate the mass of $\Xi_c(1/2^+)$ as
\begin{eqnarray}
&&M(\Xi_c,1/2^+) \big\vert_{\rm theo} \nonumber\\
&&= M(\Lambda_c,1/2^+)+ (M_{1,2}(0^+)- M_3(0^+))
= 2467 \, \mbox{MeV},\nonumber
\end{eqnarray}
which beautifully agrees with the experimental value given in Eq.~(\ref{Xi mass}).
Similarly, the masses of $\Lambda_c(1/2^-)$ and $\Xi_c(1/2^-)$ are predicted as
\begin{eqnarray}
&& M( \Lambda_c ,1/2^-)  \big\vert_{\rm theo}  \nonumber\\
&&= M(\Lambda_c,1/2^+)+ (M_{3}(0^-)- M_3(0^+))
=2826 \mbox{ MeV,} \nonumber\\
&& M(\Xi_c,1/2^-) \big\vert_{\rm theo} \nonumber\\
&&= M(\Lambda_c,1/2^+)+ (M_{1,2}(0^-)- M_3(0^+))
=2704 \mbox{ MeV.}\nonumber
\end{eqnarray}

Note that the above results are independent from the choice of $A$.
We, however, have to fix $A$ to determine the Lagrangian parameters $m_1^2$ and $m_2^2$.
By assuming $A=5/3$, 
\begin{eqnarray}
&& m_0^2 = (1031 \mbox{ MeV})^2, \nonumber\\
&& m_1^2 = (606.3 \mbox{ MeV})^2,  \\
&& m_2^2 = -(274.4 \mbox{ MeV})^2. \nonumber
\end{eqnarray}
It should be noted here that the value of $m_2^2$ happens to be negative, but it is 
perfectly all right because this is a parameter of the Lagrangian whose sign is not
constrained.

\subsection{Method II}

As an alternative, we may use the charmed baryon masses to determine the mass differences
of the diquarks.
In order to determine the diquark masses, we need the masses of $M(\Lambda_c,1/2^+)$,
$M(\Xi_c, 1/2^+)$ and $M(\Lambda_c, 1/2^-)$. The first two are experimentally given as~\cite{PDGonline}
\begin{eqnarray}
&& M(\Lambda_c,1/2^+) = 2286.46 \hbox{ MeV,}\label{Lambda c mass}\\
%&& M (\Xi^+_c ) = 2467.93 {\rm MeV},  M (\Xi^0_c) = 2470.91 {\rm MeV}\\
&& M(\Xi_c,1/2^+)=\frac{1}{2} (M(\Xi_c^+)+M(\Xi_c^0)) =2469.42\hbox{ MeV.}\nonumber\\
&&\label{Xi mass}
\end{eqnarray}
However, the masses of the negative-parity states have not been determined by experiment.
$\Lambda_c (1/2^-)$ observed at 2592 MeV is not a $0^-$ diquark bound state, but is rather 
a $P$-wave bound state of the $0^+$ diquark (see discussions in Sec.~\ref{Discussions}).
Then the bound state of a $0^-$ diquark and a charm quark is the second ($\rho$-mode) $1/2^-$ state and we do not have experimental data yet. 
Therefore we here use a quark model prediction of
the second $\Lambda_c(1/2^-)$ from Ref.~\cite{Yoshida:2015tia},
\begin{eqnarray}
&& M(\Lambda_c,1/2^-) =2890 \hbox{ MeV}.\label{Lambda^-mass}
\end{eqnarray}
Using these data, we find
\begin{eqnarray}
&& M_1(0^+) - M_3(0^+) \nonumber\\
&& = M(\Xi_c, 1/2^+) - M(\Lambda_c,1/2^+)= 183 \hbox{ MeV}.
\label{M13massdiff}\\
&& M_3(0^-) - M_3(0^+) \nonumber\\
&& = M(\Lambda_c, 1/2^-) - M(\Lambda_c,1/2^+)= 604 \hbox{ MeV}.
\label{M+-massdiff}
\end{eqnarray}
By using the lattice data for the lightest diquark mass as an input,
\begin{eqnarray}
&& M_3(0^+) = 725 \mbox{ MeV} ,
\end{eqnarray}
we obtain, from Eqs.~(\ref{M13massdiff}), (\ref{M+-massdiff}) and (\ref{rel-diff}),
\begin{eqnarray}
&& M_1(0^+) = 906 \mbox{ MeV},\\
&& M_3(0^-) = 1329 \mbox{ MeV},\\
&& M_1(0^-) = 1212 \mbox{ MeV}.
\end{eqnarray}
Then the masses of the $1/2^-$ charmed baryons are predicted as
\begin{eqnarray}
&& M(\Xi_c,1/2^-) \big\vert_{\rm theo} = 2772\, \mbox{MeV} \ .
\end{eqnarray}

Again, the above results are independent from the choice of $A$, while we can determine the parameters of the Lagrangian for the Method II, by setting $A=5/3$, as
\begin{eqnarray}
&& m_0^2 = (1070\mbox{ MeV})^2, \nonumber\\
&& m_1^2 = (632 \mbox{ MeV})^2, \\
&& m_2^2 = -(213\mbox{ MeV})^2. \nonumber
\end{eqnarray}

\subsection{Discussions}\label{Discussions}

\begin{table}[t]
\begin{center}
\begin{tabular}{l|r|r|r}
& Method I & Method II & Experiment \\
\hline\hline
$M_3(0^+)$ (MeV)　&725* & 725* \\
$M_{1,2}(0^+)$ (MeV)&906*& 906 \\
$M_3(0^-)$ (MeV)&1265*& 1329\\
$M_{1,2}(0^-)$ (MeV)&1142 & 1212\\
\hline
$M(\Lambda_c, 1/2^+)$(MeV) &2286* &2286* &2286.46 \\
$M(\Xi_c, 1/2^+)$(MeV) & 2467& 2469* & 2469.42\\
$M(\Lambda_c,1/2^-)$(MeV) & 2826& 2890* & 2592 \\
$M(\Xi_c,  1/2^-)$(MeV) & 2704& 2772 & 2793 \\
\hline
$m_0^2$ (MeV$^2$)　& (1031)$^2$ & (1070)$^2$\\
$m_1^2$ (MeV$^2$)　　& (606)$^2$ &　(631)$^2$\\
$m_2^2$ (MeV$^2$)　& $-(274)^2$ & $-(210)^2$\\
\hline\hline
\end{tabular}
\end{center} 
\caption{Parameters of the chiral effective theory and the predicted diquark and baryon masses. The asterisk is for the input values. The experimental value of the $\Xi_c$ mass is the (charge) average of $\Xi_c^0$ and $\Xi_c^+$.}
\label{tab:effective_theory}
\end{table}%

The results obtained from the two methods are summarized in Table \ref{tab:effective_theory}.
One immediately sees that the two methods give almost identical results. This simply indicates that our scheme works very well with the diquark masses given by the lattice QCD calculation.

A prominent feature of the mass spectrum is the inverse ordering of $\Lambda_c(1/2^-)$ and 
$\Xi_c(1/2^-)$. This is anomalous from the quark model viewpoint because $\Xi_c =(csq)$
contains a strange quark and
is expected to be heavier than $\Lambda_c(cqq)$ for the same quantum numbers.
A naive estimate would conclude $M(\Xi_c) \sim M(\Lambda_c)+ 200$ (MeV), while 
the present chiral dynamics predicts $M(\Xi_c) \sim M(\Lambda_c)-120$ (MeV) for the $1/2^-$ states.
The difference comes from the combination of the $U_A(1)$ anomaly term and the second-order 
chiral-symmetry breaking term as is seen in Eq.~(\ref{rel-diff}).

The PDG~\cite{PDGonline} reports a $\Lambda_c(1/2^-)$ state at 2592 MeV and a $\Xi_c(1/2^-)$ state at 2793 MeV (Table \ref{tab:effective_theory}). 
However, they may not directly be compared with our predictions.
There are two competing structures for the negative parity $1/2^-$ baryon resonances, either a bound state of
$0^+$ diquark and a charm in $P$ wave ($L=1$) ($\lambda$-mode), or a bound state of $0^-$ and a charm in $S$ wave ($\rho$-mode).
Our diquark picture assumes the $\rho$-mode excited states, where the diquark itself is
excited.
In the quark model analysis, the $\rho$-mode states are in general heavier than
the $\lambda$-mode states~\cite{CopleyPRD20}.
In fact, $\Lambda_c (2592)$ fits to the $\lambda$-mode in the quark
model very well~\cite{Yoshida:2015tia}.

On the other hand, because of the inverse ordering, our prediction of $\Xi_c(1/2^-)$
comes as low as the observed state, $\Xi_c (2793)$, 
while, in the quark model, $\Xi_c (2793)$ would be assigned to the $\lambda$-mode
excitation.
It is interesting to see whether $\Xi_c (2793)$ is possible to be the $\rho$-mode excitation.
If so, we expect to have two $\Xi_c(1/2^-)$ states in the same energy region.

In the present numerical analysis, the parameter $A$ is fixed to $5/3\sim1.67$. 
This value comes from the conventional wisdom in the quark model 
that the ratio of the constituent quark masses of $s$ and $u/d$ is given by
\begin{eqnarray}
&& A=\frac{{\cal M}_{\rm eff}(s)}{{\cal M}_{\rm eff}(u/d)} \sim \frac{5}{3}.
\end{eqnarray}
Let us estimate $A$ according to the definition, Eq.~(\ref{Apara}).
First, $f_s=128$ MeV, and $f_{\pi}=92$ MeV are determined from the weak decays of the pseudoscalar mesons.
$g_s$ is the coupling constant of the pion to the $u, d$ quark. 
It can be related to the $\pi NN$ coupling constant, {\i.e.,}
$g_s= \frac{1}{3} g_{\pi NN} \sim 4.2$,
Here we use $g_{\pi NN} =12.5$, which is determined from the Goldberger-Treiman relation.
Finally $m_s$ is the current strange quark mass determined in the chiral perturbation theory,
$m_s\sim 85-105$ MeV~\cite{PDGonline}.
From these values, we obtain $A \sim 1.61-1.67$, which agrees with our choice.

While this estimate is plausible, we check how the results depend on the value of $A$.
We recalculate the Lagrangian parameters for $A=1.5$ and 2 for the Method II.
Noting that $m_0^2$ does not depend on $A$, we obtain
\begin{eqnarray}
&& m_1^2 = ( 696\mbox{ MeV})^2,\nonumber\\
&& m_2^2 = -( 328\mbox{ MeV})^2, 
\end{eqnarray}
for $A=1.5$ and 
\begin{eqnarray}
&& m_1^2 = (552\mbox{ MeV})^2,\nonumber\\
&& m_2^2 = (96\mbox{ MeV})^2, 
\end{eqnarray}
for $A=2$.

\begin{table}[tpb]
\begin{center}
\begin{tabular}{c|r|r }
　$A$　 & $m_1^2$ (MeV$^2$) & $m_2^2$ (MeV$^2$) \\
\hline
$1.50$ & (696)$^2$ & $-(328)^2$\\
$1.67$ & (631)$^2$ & $-(210)^2$\\
$1.91$ & (569)$^2$ & $0$\\
$2.00$ & $(552)^ 2$ & $(96)^2$\\
\end{tabular}
\end{center} 
\caption{$A$ dependences of the parameters, $m_1^2$ and $m_2^2$ for the Method II.}
\label{tab:A-dependence}
\end{table}
We summarize the $A$ dependence in Table~\ref{tab:A-dependence}.
It is found that the value of $m_2^2$ is sensitive to the choice of $A$.
In fact, we can fit to the charmed baryon masses without $m_2^2$ term 
for $A=1.91$. On the other hand,  the $U_A(1)$ anomaly term, $m_1^2$, is more stable 
and is consistently dominant.

\section{Conclusion}

In this paper, we have proposed a chiral effective theory of scalar and pseudoscalar diquarks.
Based on the linear representations, we find 
that the color $\bar 3$, flavor $\bar 3$ and spin-parity $0^+$ diquark, $S$, and 
the $0^-$ diquark, $P$, with the same color and flavor, form a chiral $(\bar 3, 1) + (1, \bar 3)$
representation. Their mass difference comes from spontaneous chiral symmetry breaking (SCSB). 

A linear-sigma-model Lagrangian is constructed with three mass parameters, $m_0^2$,
$m_1^2$ and $m_2^2$.
Among them, $m_0^2$ represents the chiral invariant mass.
On the other hand, the $m_1^2$ and $m_2^2$ terms yield masses under SCSB. 
Furthermore, the $m_1^2$ term breaks the axial $U_A(1)$ symmetry and thus represents 
the $U_A(1)$ anomaly.
It is shown that the $m_0^2$ mass is diagonal in the chirality of the diquark, while 
the $m_1^2$ and $m_2^2$ masses are off-diagonal, connecting the left and right diquarks.
The scalar and pseudoscalar diquarks are mass eigenstates and their
mass difference is given by $m_1^2$ and $m_2^2$.
We also find that the coupling of the pseudoscalar octet mesons to the diquarks 
satisfies the generalized Goldberger-Trieman relation.

By introducing the finite quark mass effects, we find that the contributions of the $U_A(1)$ anomaly 
mass $m_1^2$ depend on the quark mass in a reversed manner compared 
to the $m_2^2$ contributions. As a result, we find the inverse mass ordering of the
negative-parity diquarks, $M(us/ds, 0^-) < M(ud, 0^-)$.

In order to estimate the coupling constants in the effective Lagrangian, we
take into account the results of lattice QCD calculations of diquark masses and also 
the masses of the bound states of a charm quark and a diquark, {\it i.e.,}
singly charmed baryons. We propose two methods of determining the
parameters, which give similar results.
The most prominent feature of the diquark picture of the charmed baryon
is the reversed ordering of $\Lambda_c(1/2^-)$ and $\Xi_c(1/2^-)$.
We predict a lower mass for $\Xi_c(1/2^-)$.
This inversion is caused by the $U_A(1)$ anomaly term.
A similar mass inversion was seen also in the scalar meson spectrum in a chiral
effective theory approach~\cite{Kuroda:2019jzm}.

Numerical values of the parameters in the effective Lagrangian
show that the $U_A(1)$ anomaly term is dominant for the mass
difference between the positive- and negative-parity diquarks.

So far, we have introduced only the scalar and pseudoscalar diquarks. 
It is interesting to extend
this approach to vector and axial-vector diquarks.
Considering finite temperature and baryon density is another
direction to explore, as the diquark masses might change due to
restoration of the chiral-symmetry breaking.
These are subjects of future studies.
%

%
% @@@ =================================================================
%
\section*{Acknowledgments}
We thank Dr. Daisuke Jido for useful discussions. This work was supported in part by JSPS KAKENHI Grant Nos. JP16K05345 (M.H.), JP17K14277 (K.S.), JP19H05159 (M.O.), and 
also by NNSFC (No. 11775132) (Y.R.L).
%
%
% @@ ==================================================================
%

\end  {document}